\documentclass[aps,prd,preprint,eqsecnum,floats,
showpacs,superscriptaddress]{revtex4-1}
\usepackage{amssymb} 
\usepackage{amsmath}
\usepackage{hyperref} 
\usepackage{graphicx}
\usepackage{subfigure}

\title{Cosmic Axion Bose-Einstein Condensation}
\begin{document}

\author{Nilanjan Banik }
\affiliation{University of Florida}

\author{Pierre Sikivie}
\affiliation{University of Florida}

\begin{abstract}

QCD axions are a well-motivated candidate for cold dark matter. Cold axions 
are produced in the early universe by vacuum realignment, axion string decay 
and axion domain wall decay.  We show that cold axions thermalize via their 
gravitational self-interactions, and form a Bose-Einstein condensate.  As a 
result, axion dark matter behaves differently from the other proposed forms 
of dark matter.  The differences are observable.
 
\end{abstract}

\maketitle

\section{QCD Axions}

The theory of strong interactions, called "quantum chromodynamics" 
or QCD for short, has in its Lagrangian density a "$\theta$-term" 
\cite{tHooft1,tHooft2,Jackiw,Callan}
\begin{equation} 
\mathcal{L}_{\theta}=
\theta\frac{g_{s}^{2}}{32\pi^{2}}\tilde{G}^{a\mu\nu}G^{a}_{\mu\nu} 
\end{equation} 
where $\theta$ is an angle between 0 and $2\pi$, $g_{s}$ is the 
coupling constant for strong interactions, and $G_{\mu\nu}^a$ is 
the gluon field tensor. The $\theta$-term is a 4-divergence and 
therefore has no effects in perturbation theory.  However, it can 
be shown to have non-perturbative effects, and these are important 
at low energies/long distances. Since $\mathcal{L}_\theta$ is P and 
CP odd, QCD violates those discrete symmetries when $\theta \neq 0$.  
The strong interactions are observed to be P and CP symmetric, and 
therefore $\theta$ must be small.  The experimental upper bound on 
the electric dipole moment of the neutron implies 
$\theta \lesssim 0.7\times 10^{-11}$ \cite{Baker,Gpkim}.
In the Standard Model of particle physics there is no reason for 
$\theta$ to be small; it is expected to be of order one.  That 
$\theta$ is less than $10^{-11}$ is a puzzle, referred to as the 
strong CP problem.

Peccei and Quinn proposed \cite{PQ, PQ2} solving the strong CP 
problem by introducing a global $U(1)_{\rm PQ}$ symmetry which 
is spontaneously broken.  When some conditions are met, the 
parameter $\theta $ is promoted to a dynamical field 
$\frac{\phi(x)}{f_{a}}$, where $f_{a}$ is the energy scale at 
which $U(1)_{\rm PQ}$ is spontaneously broken, and $\phi(x)$ 
the associated Nambu-Goldstone boson field.  The theory now 
depends on the expectation value of $\phi(x)$.  The latter 
minimizes the QCD effective potential.  It can be shown 
that the minima of the QCD effective potential occur where 
$\theta=0$ \cite{Vafawitten}.  The strong CP problem is 
thus solved if there is a Peccei-Quinn symmetry.

Axions are the quanta of the field $\phi(x)$ 
\cite{Weinbergaxion,Wilczek}. Axions acquire mass due to 
the non-perturbative effects that make QCD depend on $\theta$.
The axion mass is given by 
\begin{equation}
m_{a}\simeq 10^{-6}\mathrm{eV}
\left(\frac{10^{12}\hspace{1mm}\mathrm{GeV}}{f_{a}}\right)
\label{amass}
\end{equation}
when the temperature is zero.

\section{Production of cold axions}

The equation of motion for $\phi(x)$ is 
\begin{equation}
D_{\mu}D^{\mu}\phi(x)+V_{a}'(\phi(x))=0
\end{equation}
where $V_a^\prime$ is the derivative of the effective potential 
with respect to the axion field and $D_{\mu}$ is the covariant 
derivative with respect to space-time coordinates.  The effective 
potential may be written
\begin{equation}
V_{a}= m_{a}(t)^{2}f_{a}^{2}
\left[1-\cos\left(\frac{\phi(x)}{f_{a}}\right)\right].
\end{equation} 
The axion mass is temperature- and hence time-dependent.  It reaches 
its zero-temperature value, Eq. (\ref{amass}), at temperatures well 
below 1 GeV.  At temperatures much larger than 1 GeV, $m_a$ is 
practically zero.  The axion field starts to oscillate  
\cite{Preskill,Abbottsikivie,Dine} at a time 
$t_1$ after the Big Bang given by 
\begin{equation}
m(t_{1})\cdot t_{1}= 1~~\ .
\label{t1}
\end{equation} 
Throughout we use units in which $\hbar = c =1$.  $t_{1}$ is 
approximately $2\times 10^{-7}\mathrm{s}
\left(\frac{f_{a}}{10^{12}\hspace{1mm}\mathrm{GeV}}\right)^{1/3}$.
The temperature of the primordial plasma at that time is 
$T_{1}\simeq 1\hspace{1mm}\mathrm{GeV}
\left(\frac{10^{12}\hspace{1mm}\mathrm{GeV}}{f_{a}}\right)^{1/6}$.  
The $\phi(x)$ oscillations describe a population of axions called 
``of vacuum realignment".  Their momenta are of order $t_1^{-1}$ 
at time $t_1$, and are red-shifted by the expansion of the universe 
after $t_1$:
\begin{equation}
\delta p(t) \sim {1 \over t_1}{R(t_1) \over R(t)}
\label{mom}
\end{equation}
where $R(t)$ is the scale factor.  As a result the axions are 
non-relativistic soon after $t_1$, and today they are extremely 
cold.  The fact that they are naturally abundant, weakly coupled
and very cold, and that they solve the strong CP problem as well,
makes axions an attractive candidate for the dark matter of the 
universe.

The number of axions produced depends on various circumstances, 
in particular whether inflation occurred before or after the  
phase transition in which $U(1)_{\rm PQ}$ is spontaneously 
broken, hereafter called the PQ phase transition.  For a review, 
see ref. \cite{axcosmo}.  If inflation occurs after, it homogenizes 
the axion field within the observable universe.  The initial value 
of the axion field may then be accidentally close to the CP conserving 
value in which case the cold axion population from vacuum realignment 
is suppressed.  If inflation occurs before the PQ phase transition, 
there is always a vacuum realignment contribution (because the axion 
field has random unrelated values in different QCD horizons) and there 
are additional contributions from axion string decay and axion domain 
wall decay.  The number density of cold axions is  
\begin{equation}
n(t)\simeq \frac{4\times 10^{47}}{\mathrm{cm}^{3}}X
\left(\frac{f_{a}}{10^{12}\mathrm{GeV}}\right)^{5/3}
\left(\frac{R(t_{1})}{R(t)}\right)^{3}
\label{aden}
\end{equation}  
where $X$ is a fudge factor.  If inflation occurs before the PQ phase 
transition, $X$ is of order 2 or 20 depending on whose estimate of the 
string decay contribution one believes.  If inflation occurs after the 
PQ phase transition, $X$ is of order $\frac{1}{2}\sin^{2}\alpha_{1}$, 
where $\alpha_{1} = \phi(t_1)/f_a$ is the initial misalignment angle.

Cold axions are effectively stable because their lifetime is 
vastly longer than the age of the universe.  The number of axions 
is effectively conserved.  The phase-space density of cold axions 
implied by Eqs.~(\ref{mom}) and (\ref{aden}) is \cite{PSyangBEC}
\begin{equation}
\mathcal{N}\sim n \frac{(2\pi)^{3}}{\frac{4\pi}{3}(m\delta v)^{3}}
\sim 10^{61}X
\left(\frac{f_{a}}{10^{12}\hspace{1mm}\mathrm{GeV}}\right)^{8/3}.
\label{phsp}
\end{equation}  
$\mathcal{N}$ is the average occupation number of those axion states 
that are occupied.   Because their phase-space density is huge and 
their number is conserved, cold axions may form a Bose-Einstein 
condensate (BEC).  The remaining necessary and sufficient condition 
for the axions to form a BEC is that they thermalize.  Assuming
thermal equilibrium, the critical temperature is 
\cite{PSyangBEC,CosAxTher} 
\begin{equation}
T_{c}(t)=
\left(\frac{\pi^{2}n(t)}{\zeta(3)}\right)^{1/3}
\simeq 300 \hspace{3mm}\mathrm{GeV}~X^{1/3}
\left(\frac{f_{a}}{10^{12}\hspace{1mm}\mathrm{GeV}}\right)^{5/9}~~
\frac{R(t_{1})}{R(t)}~~~\ .
\label{TC}
\end{equation}
The critical temperature is enormous because the cosmic axion density 
is so very high.   The formula given in Eq.~(\ref{TC}) differs from 
the one for atoms because, in thermal equilibrium,  most of the 
non-condensate axions would be relativistic.  

The question is whether the axions thermalize.  This is not at all 
obvious since axions are extremely weakly coupled.  Note that for 
Bose-Einstein condensation to occur, it is not necessary that full 
thermal equilibrium be reached. It is sufficient that the rate of 
condensation into the lowest energy available state be larger than 
the inverse age of the universe.  Whether this happens is the issue 
which we address next.

\section{Axion-axion interactions}

Axions interact by $\lambda \phi^4$ self-interactions and by 
gravitational self-interactions. In this section we discuss 
these two processes in detail and calculate the corresponding 
relaxation rates \cite{CosAxTher}.  Let us introduce a cubic 
box of volume $V=L^3$, with periodic boundary conditions at 
the surface. The axion field and its canonically conjugate 
field $\pi(\vec{x},t)$ may be written as
\begin{eqnarray}
\phi(\vec{x},t)&=&\sum\limits_{\vec{n}}
(a_{\vec{n}}(t)\Phi_{\vec{n}}(\vec{x})+ 
a^{\dagger}_{\vec{n}}(t)\Phi^{*}_{\vec{n}}(\vec{x})) \\
\pi(\vec{x},t)&=&\sum\limits_{\vec{n}}
(-i\omega_{\vec{n}})(a_{\vec{n}}(t)\Phi_{\vec{n}}(\vec{x})
- a^{\dagger}_{\vec{n}}(t)\Phi^{*}_{\vec{n}}(\vec{x}))
\end{eqnarray}
inside the box, where 
\begin{equation}
\Phi_{\vec{n}} (\vec{x}) = 
\frac{e^{i\vec{p}_{\vec{n}}\cdot\vec{x}}}{\sqrt{2\omega_{\vec{n}}V}}~~\ ,
\end{equation}
$\vec{p}_{\vec{n}}=2\pi\vec{n}/L$, $\vec{n}=(n_{1},n_{2},n_{3})$ where 
$n_{1}$, $n_{2}$, $n_{3}$ are integers, and 
$\omega_{\vec{n}}=\sqrt{p_{\vec{n}}^2 + m^2}$. The creation and 
annihilation operators satisfy canonical equal-time commutation 
relations
\begin{equation}
[a_{\vec{n}}(t),a^{\dagger}_{\vec{n}'}(t)]=\delta_{\vec{n},\vec{n}'},  
\hspace{2cm} [a_{\vec{n}}(t),a_{\vec{n}'}(t)]=0.
\end{equation}
The Hamiltonian, including $\lambda\phi^4$ self-interactions, is 
\begin{equation}
H= \sum\limits_{\vec{n}}\omega_{\vec{n}}a_{\vec{n}}^{\dagger}a_{\vec{n}} 
\hspace{3mm}+ 
\sum\limits_{\vec{n}_{1},\vec{n}_{2},\vec{n}_{3},\vec{n}_{4}} 
\frac{1}{4}\Lambda_{s \hspace{0.18cm}    
\vec{n}_{1},\vec{n}_{2}}^{\vec{n}_{3}\vec{n}_{4}} 
a_{\vec{n}_{1}}^{\dagger}a_{\vec{n}_{2}}^{\dagger}a_{\vec{n}_{3}}a_{\vec{n}_{4}}
\label{lambda}
\end{equation}
where 
\begin{equation}
\Lambda_{s \hspace{0.18cm}\vec{n}_{1},\vec{n}_{2}}^{\vec{n}_{3}\vec{n}_{4}} = 
\frac{-\lambda}{4m^2 V}\delta_{\vec{n}_{1}+\vec{n}_{2},\vec{n}_{3}+\vec{n}_{4}}
~~\ .
\label{Ls}
\end{equation}
The Kronecker-delta ensures 3-momentum conservation. When deriving Eq. \
(\ref{lambda}), axion number violating terms such as $aaaa$, 
$a^{\dagger}a^{\dagger}a^{\dagger}a^{\dagger}$, $a^{\dagger}aaa$, 
$a^{\dagger}a^{\dagger}a^{\dagger}a$ are neglected.  Indeed, in 
lowest order they allow only processes that are forbidden by 
energy-momentum conservation. In higher orders they do lead to 
axion number violating processes but only on times scales that 
are vastly longer than the age of the universe.

The gravitational self-interactions of the axion fluid are 
described by Newtonian gravity since we only consider interactions 
on sub-horizon scales. The interaction Hamiltonian is 
\begin{equation}
H_{g} = -\frac{G}{2}
\int d^{3}x\hspace{1mm}d^{3}x'\hspace{1mm} 
\frac{\rho(\vec{x},t)\rho(\vec{x}',t)}{|\vec{x}-\vec{x}'|}
\end{equation}  
where $\rho(\vec{x},t)=\frac{1}{2}(\pi^{2}+m^2\phi^{2})$ is the 
axion energy density.  In terms of creation and annihilation 
operators \cite{CosAxTher} 
\begin{equation}
H_{g}= \sum\limits_{\vec{n}_{1},\vec{n}_{2},\vec{n}_{3},\vec{n}_{4}} 
\frac{1}{4}\Lambda_{g \hspace{0.18cm}\vec{n}_{1},\vec{n}_{2}}
^{\vec{n}_{3}\vec{n}_{4}} a_{\vec{n}_{1}}^{\dagger}
a_{\vec{n}_{2}}^{\dagger}a_{\vec{n}_{3}}a_{\vec{n}_{4}}
\end{equation} 	
where 
\begin{equation}
\Lambda_{g \hspace{0.18cm}\vec{n}_{1},\vec{n}_{2}}^{\vec{n}_{3}\vec{n}_{4}} 
= -\frac{4\pi Gm^2}{V}\delta_{\vec{n}_{1}+\vec{n}_{2},\vec{n}_{3}+\vec{n}_{4}}
\left(\frac{1}{|\vec{p}_{\vec{n}_{1}}-\vec{p}_{\vec{n}_{3}}|^2} + 
\frac{1}{|\vec{p}_{\vec{n}_{1}}-\vec{p}_{\vec{n}_{4}}|^2}\right).
\label{Lg}
\end{equation}	
$H_g$ must be added to the RHS of Eq.~(\ref{lambda}).  

In summary, we have found that the axion fluid is described by a set 
of coupled quantum harmonic oscillators. We now estimate the resulting
relaxation rates.  There are two different regimes of relaxation depending 
on the relative values of the relaxation rate $\Gamma$ and the energy 
dispersion $\delta \omega$.  The condition $\Gamma << \delta \omega$ 
defines the ``particle kinetic regime", whereas $\Gamma >> \delta \omega$ 
defines the ``condensed regime".  Most physical systems relax in the 
particle kinetic regime.  Axions on the other hand relax in the 
condensed regime.

\subsection{Particle kinetic regime}

When $\Gamma << \delta \omega$, the rate of change of the occupation 
numbers ${\mathcal N}_i$ ($i = 1, 2, .. M$) of $M$ coupled oscillators 
is given by 
\begin{multline}
\left\langle\dot{\mathcal{N}}_{l}\right\rangle= 
\sum\limits_{i,j,k=1}^{M}\frac{1}{2}|\Lambda_{ij}^{kl}|^{2}
[\mathcal{N}_{i}\mathcal{N}_{j}(\mathcal{N}_{l}+1)(\mathcal{N}_{k}+1) \\
-(\mathcal{N}_{i}+1)(\mathcal{N}_{j}+1)\mathcal{N}_{l}\mathcal{N}_{k}]
2\pi\delta(\Omega_{ij}^{lk})~~+~~{\mathcal O}(\Lambda^3)
\label{pkr}
\end{multline}
where $\Omega_{ij}^{kl} = \omega_k + \omega_l - \omega_i - \omega_j$, 
and the $\Lambda_{ij}^{kl}$ are the relevant couplings, such as are
given in Eqs.~(\ref{Ls}) and (\ref{Lg}) for axions.  If we substitute 
the couplings due to $\lambda \phi^4$ interactions, Eq.~(\ref{Ls}),
and replace the sums over modes by integrals over momenta, we obtain
\cite{Semikoz,CosAxTher} 
\begin{multline}
\left\langle\dot{\mathcal{N}_1}\right\rangle = 
\frac{1}{2\omega_{1}}\int
\frac{d^{3}p_{2}}{(2\pi)^{3}2\omega_{2}}
\frac{d^{3}p_{3}}{(2\pi)^{3}2\omega_{3}}
\frac{d^{3}p_{4}}{(2\pi)^{3}2\omega_{4}}
\lambda^{2}(2\pi)^{4}\delta^{4}(p_{1}+p_{2}-p_{3}-p_{4})	\\
 \times\frac{1}{2}
[(\mathcal{N}_{1} + 1)(\mathcal{N}_{2} + 1) \mathcal{N}_{3} \mathcal{N}_{4}
- \mathcal{N}_{1} \mathcal{N}_{2} (\mathcal{N}_{3} + 1)(\mathcal{N}_{4} + 1)]
~~\ ,
\label{Boltz}
\end{multline}
where ${\cal N}_1 \equiv {\cal N}_{\vec{p}_1}$ and so forth.
When the states are not highly occupied (${\mathcal N} \lesssim 1$),
Eq.~(\ref{Boltz}) implies the standard formula for the relaxation rate
\begin{equation}
\Gamma \sim \frac{\dot{\mathcal{N}}}{\cal N}\sim n\sigma\delta v
\label{rel1}
\end{equation}
where $\sigma=\lambda^{2}/64\pi m^{2}$ is the scattering cross-section 
due to $\lambda \phi^4$ interactions, $n$ is the particle density and 
$\delta v$ is the velocity dispersion.  On the other hand, when the 
states are highly occupied (${\mathcal N} >> 1$), Eq.~(\ref{Boltz}) 
implies 
\begin{equation}
\Gamma \sim n \sigma \delta v \mathcal{N}.
\label{rel2}
\end{equation}
The relaxation rate is enhanced by the degeneracy factor, which is huge 
(${\mathcal N} \sim 10^{61}$) in the axion case.  The process of 
Bose-Einstein condensation occurs as a result of scatterings 
$a(\vec{p}_1) + a(\vec{p}_2) \leftrightarrow a(\vec{p}_3) + a(\vec{p}_4)$ 
in which ${\cal N}_1$, ${\cal N}_2$ and ${\cal N}_3$ are of order 
the large degeneracy factor ${\cal N}$ whereas ${\cal N}_4 << {\cal N}$.
Eq.~(\ref{Boltz}) implies that, as a result of such scatterings, the 
occupation number of the lowest available energy state grows 
exponentially with the rate given in Eq.~(\ref{rel2}) 
\cite{Semikoz,PSyangBEC,Jaeckel}.

In contrast to $\lambda \phi^4$ interactions, gravitational interactions 
are long-range.  The cross-section for gravitational scattering is 
infinite due to the contribution from very small angle (forward) 
scattering.  But forward scattering does not contribute to relaxation, 
whereas scattering through large angles does contribute.  (The issue does
not arise in the case of $\lambda \phi^4$ interactions, for which there is no 
peak in the differential cross-section for forward scattering and scattering 
is generically through large angles.)  The upshot is that Eqs.~(\ref{rel1}) 
and (\ref{rel2}) are still valid for estimating the relaxation rate by 
gravitational interactions in the particle kinetic regime provided one 
uses for $\sigma$ the cross-section for large angle scattering.  That 
cross-section is finite and equals
\begin{equation}
\sigma_{g} \sim \frac{4G^{2}m^{2}}{(\delta v)^4}
\end{equation}
in order of magnitude.

\subsection{Condensed regime}
 
When $\Gamma >> \delta \omega$, one cannot use Eq.~(\ref{pkr}) because
the derivation of that equation involves an averaging over time that 
is valid only when $\Gamma << \delta \omega$.  Instead we will use
the equations
\begin{equation}
i \dot{a}_l(t) = \omega_l a_l(t) + \sum\limits_{i,j,k=1}^M
{1 \over 2} \Lambda_{kl}^{ij} a_k^\dagger a_i a_j 
\label{Heis1}
\end{equation}
which follow directly from the Hamiltonian, Eq.~(\ref{lambda}).  It 
is convenient to define $c_{l}(t) \equiv a_{l}(t)e^{i\omega_{l}t}$, 
in terms of which Eq. (\ref{Heis1}) becomes 
\begin{equation}
\dot{c}_{l}(t)=-i\sum\limits_{i,j,k=1}^{M}\frac{1}{2}\Lambda_{kl}^{ij}
c_{k}^{\dagger}c_{i}c_{j}e^{i\Omega_{ij}^{kl}t}
\label{Heis2} 
\end{equation}
where $\Omega_{ij}^{kl} \equiv \omega_k + \omega_l - \omega_i - \omega_j$, 
as before.  Further, because the occupation numbers of the occupied states 
are huge, we write $c_{l}$ as a sum of a classical part $C_{l}$ and a 
quantum part $d_{l}$ 
\begin{equation}
c_{l}(t)=C_{l}(t)+d_{l}(t)~~\ .
\end{equation} 
The $C_{l}$ are c-number functions of order $\sqrt{\mathcal{N}_l}$ describing 
the bulk of the axion fluid. They satisfy the equations of motion 
\begin{equation}
\dot{C}_{l}(t)=-i\sum\limits_{i,j,k = 1}^{M}\frac{1}{2}
\Lambda_{kl}^{ij}C_{k}^{*}C_{i}C_{j}e^{i\Omega_{ij}^{kl}t}.
\label{class}
\end{equation}
The $d_{l}$ and $d_l^\dagger$ are annihilation and creation operators 
satisfying canonical commutation relations.  Quantum statistics plays 
the essential role in determining the {\it outcome} of relaxation to 
be the Bose-Einstein distribution.  However, we may use classical physics 
to estimate the {\it rate} of relaxation.  The relaxation rate is the 
inverse time scale over which $C_{l}(t)$ changes by an amount of order 
$C_l(t)$.  

The sum in Eq.~(\ref{class}) is dominated by those states that are highly 
occupied.  Let $K$ be the number of such states.  Using the fact that in the 
condensed regime $\Omega_{ij}^{kl} t << 1$, we may rewrite Eq.~(\ref{class}) 
as
\begin{equation}
\dot{C}_{l}(t)\sim -i \sum\limits_{i,j,k = 1}^{K}\frac{1}{2}
\Lambda_{kl}^{ij} C_{k}^{*}C_{i}C_{j}~~\ .
\end{equation}
If we substitute Eq.~(\ref{Ls}) for $\lambda\phi^{4}$ interactions, 
we get 
\begin{equation}
\dot{C}_{\vec{p}_{1}}(t)\sim i\frac{\lambda}{4m^{2}V}
\sum\limits_{\vec{p}_{2},\vec{p}_{3}}\frac{1}{2}
C_{\vec{p}_{2}}^{*}C_{\vec{p}_{3}}C_{\vec{p}_{4}}
\end{equation}
where $\vec{p}_{4}=\vec{p}_{1}+\vec{p}_{2}-\vec{p}_{3}$ and the sum 
is restricted to the highly occupied states. The sum is similar to a 
random walk with each step of order $\sim \mathcal{N}^{3/2}$ and the
number of steps of order $K^{2}$. Hence 
\begin{equation}
\dot{C}_{\vec{p}}\sim \frac{\lambda}{4m^{2}V}K\mathcal{N}^{3/2}
\sim \frac{\lambda}{4m^{2}V} N\mathcal{N}^{1/2}
\end{equation}
where we used $K \sim N/{\mathcal N}$.  Since $C_l \sim \sqrt{\cal N}$, 
the relaxation rate due to $\lambda\phi^{4}$ interactions in the condensed 
regime is \cite{PSyangBEC,CosAxTher}
\begin{equation}
\Gamma_{\lambda}\sim \frac{1}{4}n\lambda m^{-2}
\label{relsc}
\end{equation}
where $n=N/V$ is the number density of the particles in highly 
occupied states. Likewise the relaxation rate for gravitational 
scattering is found to be 
\begin{equation}
\Gamma_{g}\sim 4\pi Gnm^{2}\ell^{2}
\label{relgc}
\end{equation}
where $\ell=1/\delta p$ is the correlation length of the particles. 

The expressions estimating the relaxation rates in the condensed 
regime, Eqs.~(\ref{relsc}) and (\ref{relgc}), are very different 
from the expression, Eq. (\ref{rel2}), in the particle kinetic 
regime.  In particular, in the condensed regime, the relaxation 
rate is first order in the coupling, whereas it is second order 
in the particle kinetic regime.  But the expressions are compatible.
At the boundary between the two regimes, where $\delta \omega \sim \Gamma$, 
the two estimates agree.  At that boundary, up to factors of order 2 or so, 
\begin{equation}
\delta v \mathcal{N}\sim 
\delta v\frac{n}{(\delta p)^{3}}\sim 
\frac{n}{m^{2}\delta\omega}\sim \frac{n}{m^{2}\Gamma}~~\ .
\end{equation}
Substituting this into Eq.~(\ref{rel2}) yields Eq.~(\ref{relsc}).
Similarly for the relaxation rate due to gravitational 
self-interactions.  

\section{Axion BEC}

For a system of particles to form a BEC, four conditions
must be satisfied:
\begin{enumerate} 
\item the particles must be identical bosons, 
\item their number must be conserved, 
\item they must be degenerate, i.e. the average occupation number 
$\mathcal{N}$ of the states that they occupy should be order 1 or 
larger, 
\item they must thermalize. 
\end{enumerate} 
When the four conditions are satisfied, a macroscopically large 
fraction of the particles go to the lowest energy available state.
It may be useful to clarify the notion of {\it lowest energy available  
state} \cite{Banik}.  Thermalization involves interactions.  By lowest 
energy available state we mean the lowest energy state that can be 
reached by the thermalizing interactions.  In general the system has 
states of yet lower energy.  For example, and at the risk of stating 
the obvious, when a beaker of superfluid $^4$He is sitting on a table, 
the condensed atoms are in their lowest energy available state.  This 
is not their absolute lowest energy state since the energy of the 
condensed atoms can be lowered by placing the beaker on the floor.  
In the case of atoms, it is relatively clear what state the atoms 
condense into when BEC occurs.  The case of axions is more confusing 
because the thermalizing interactions, both gravity and the $\lambda \phi^4$ 
self-interactions, are attractive and therefore cause the system to be 
unstable.  When the system is unstable, the restriction to the lowest 
energy {\it available} state is especially crucial.  

We saw in the first two sections that, for cold dark matter axions, 
the first three conditions for BEC are manifestly satisfied. In this 
section we show that the fourth condition is satisfied as well 
\cite{PSyangBEC,CosAxTher}.  Cold axions will thermalize if their 
relaxation time $\tau$ is shorter than the age $t$ of the universe, 
or equivalently if their relaxation rate $\Gamma \equiv 1/\tau$ is 
greater than the Hubble expansion rate $H \sim 1/t$.

The cold axion energy dispersion is
\begin{equation}
\delta \omega (t) \simeq {(\delta p(t))^2 \over 2 m(t)}~~\ .
\label{bord}
\end{equation}
In view of Eqs.~(\ref{t1}) and (\ref{mom}), $\delta \omega(t_1) \sim 1/t_1$.
If axions thermalize at time $t_1$, we have $\Gamma(t_1) > 1/t_1$ and therefore
the thermalization is in the condensed regime or at the border between the 
particle kinetic and condensed regimes.  After time $t_1$, 
$\delta \omega (t) < 1/t$ since $m(t)$ increases sharply for 
a period after $t_1$ whereas $(\delta p(t))^2 \propto R(t)^{-2} \propto 1/t$,
since $R(t) \propto \sqrt{t}$ in the radiation dominated era.  So after $t_1$, 
axions can only thermalize in the condensed regime.

To see whether the axions thermalize by $\lambda \phi^4$ self-interactions
at time $t_1$, we may use either Eq.~(\ref{relsc}) or (\ref{rel2}).  Both 
estimates yield $\Gamma_\lambda(t_1) \sim H(t_1)$ indicating that the axions 
thermalize at time $t_1$ by $\lambda \phi^4$ self-interactions but only barely 
so.  After $t_1$ we must use Eq.~(\ref{relsc}).  It informs us that 
$\Gamma_\lambda (t) / H(t) \propto R(t)^{-3} t \propto t^{-{1 \over 2}}$,
i.e. that even if axions thermalize at time $t_1$ they stop doing so
shortly thereafter.  Nothing much changes as a result of this brief
epoch of thermalization since in either case, whether it occurs or
not, the correlation length 
$\ell(t) \equiv 1/\delta p(t) \sim t_1 R(t)/R(t_1)$.

To see whether the axions thermalize by gravitational self-interactions 
we use Eq.~(\ref{relgc}).  It implies                 
\begin{equation}
\Gamma_g(t) / H(t) \sim 8 \pi G n m^2 \ell^2 t
\sim 5 \cdot 10^{-7}~{R(t_1) \over R(t)} {t \over t_1} X           
\left({f_a \over 10^{12} {\rm GeV}}\right)^{2 \over 3}             
\label{GgoH}       
\end{equation}
once the axion mass has reached its zero temperature value,
shortly after $t_1$.  Gravitational self-interactions are    
too slow to cause thermalization of cold axions near the
QCD phase transition but, because
$\Gamma_g / H \propto R^{-1}(t) t \propto R(t)$, they do            
cause the cold axions to thermalize later on.  The RHS of 
Eq.~(\ref{GgoH}) reaches one at a time $t_{\rm BEC}$       
when the photon temperature is of order
\begin{equation}
T_{\rm BEC} \sim 500~{\rm eV}~X
\left({f_a \over 10^{12}~{\rm GeV}}\right)^{1 \over 2}~~\ .       
\label{Tbec}
\end{equation}
The axions thermalize then and form a BEC as a result of their
gravitational self-interactions.  The whole idea may seem far-fetched
because we are used to think that gravitational interactions among
particles are negligible.  The axion case is special, however, because
almost all particles are in a small number of states with very long
de Broglie wavelength, and gravity is long range.  

Systems dominated by gravitational self-interactions are inherently
unstable.  In this regard the axion BEC differs from the BECs that 
occur in superfluid $^4$He and dilute gases.  The axion fluid is 
subject to the Jeans gravitational instability and this is so 
whether the axion fluid is a BEC or not \cite{PSyangBEC}.  The Jeans 
instability causes density perturbations to grow at a rate of order 
the Hubble rate $H(t)$, i.e. on a time scale of order the age of the 
universe at the moment under consideration.  Each mode of the axion 
fluid is Jeans unstable.  We showed however that, after $t_{\rm BEC}$, 
the thermalization rate is faster than the Hubble rate.  The rate at 
which quanta of the axion field jump between modes is faster than 
the rate at which the Jeans instability develops.  So the modes are 
essentially frozen on the time scale over which the axions thermalize.

Finally, we comment on a misapprehension that appears in the 
literature.  The axions do not condense in the lowest momentum mode 
$\vec{p} = 0$.  Condensation into the $\vec{p} = 0$ state would mean
that the fluid becomes homogeneous and at rest.  Of course this is 
not what happens in the axion case since the axion fluid is Jeans
unstable.  Despite a common misconception, it is not a rule of BEC 
that the particles condense into the $\vec{p} = 0$ state.  The rule 
instead is that they condense into the lowest energy available state, 
as defined earlier.  Only in empty space, and only if the total linear 
momentum and the total angular momentum of the particles are zero, is 
the lowest energy state a state of zero momentum. It should be obvious 
that the particles do not condense in the $\vec{p} =0$ state if they are 
moving or rotating.  Nonetheless, Bose-Einstein condensation occurs.  

\section{Observational implications}

For a long time, it was thought that axions and the other proposed forms 
of cold  dark matter behave in the same way on astronomical scales and are 
therefore indistinguishable by observation.  Axion BEC changed that.  On 
time scales longer than their thermalization time scale $\tau$, axions 
almost all go to the lowest energy state available to them.  The other 
dark matter candidates, such as weakly interacting masssive particles 
(WIMPs) and sterile neutrinos, do not do this.  It was shown in 
Ref. \cite{PSyangBEC} that, on all scales of observational interest, 
density perturbations in axion BEC behave in exactly the same way as 
those in ordinary cold dark matter provided the density perturbations 
are within the horizon and in the linear regime.  On the other hand, 
when density perturbations enter the horizon, or in second order of 
perturbation theory, axions generally behave differently from ordinary 
cold dark matter because the axions rethermalize so that the state most 
axions are in tracks the lowest energy available state.

A distinction between axions and the other forms of cold dark matter 
arises in second order of perturbation theory, in the context of the 
tidal torquing of galactic halos.  Tidal torquing is the mechanism by 
which galaxies acquire angular momentum.  Before they fall onto a galactic 
halo, the axions thermalize sufficiently fast that the axions that are 
about to fall into a particular galactic gravitational potential well 
go to their lowest energy available state consistent with the total 
angular momentum they acquired from nearby protogalaxies through tidal 
torquing \cite{Banik}.  That state is a state of net overall rotation, 
more precisely a state of rigid rotation on the turnaround sphere.  In 
contrast, ordinary cold dark matter falls into a galactic gravitational 
potential well with an irrotational velocity field \cite{PSnatacaustics}. 
The inner caustics are different in the two cases.  In the case of net 
overall rotation, the inner caustics are rings \cite{causPS} whose 
cross-section is a section of the elliptic umbilic $D_{-4}$ catastrophe 
\cite{PSringsing}, called caustic rings for short.  If the velocity field 
of the infalling particles is irrotational, the inner caustics have a 
`tent-like' structure which is described in detail in 
ref.~{\cite{PSnatacaustics} and which is quite distinct from caustic rings.  
Evidence was found for caustic rings.  A summary of the evidence is given 
in ref.~\cite{PSduffy}.  Furthermore, it was shown in ref.~\cite{case} that 
the assumption that the dark matter is axions explains not only the existence 
of caustic rings but also their detailed properties, in particular the pattern 
of caustic ring radii and their overall size.

Vortices appear in the axion BEC as it is spun up by tidal torquing. The 
vortices in the axion BEC are attractive, unlike those in superfluid $^4$He 
and dilute gases.  Hence a large fraction of the vortices in the axion BEC join 
into a single big vortex along the rotation axis of the galaxy \cite{Banik}.  
Baryons and ordinary cold dark matter particles that may be present, such 
as WIMPs and/or sterile neurtinos, are entrained by the axion BEC and 
acquire the same velocity distribution.  The resulting baryonic angular 
momentum distribution gives a good qualitative fit \cite{Banik} to the 
angular momentum distributions observed in dwarf galaxies \cite{vandenAMD}.   
This resolves a long-standing problem with ordinary cold dark matter called 
the "galactic angular momentum problem" \cite{Navarro,Burk}.  A minimum 
fraction of cold dark matter must be axions to explain the data.  That 
fraction is of order 35\% \cite{Banik}.

%\bibliography{MyChapter_References}
\bibliography{References}
\bibliographystyle{cambridgeauthordate}

\end{document}